\documentclass[11pt,twoside]{article}

\usepackage{asp2014}

\aspSuppressVolSlug
\resetcounters

\begin{document}

\title{ADASS 2024 - Stereograph: stereoscopic event reconstruction using graph neural networks applied to CTAO}

\author{Hana Ali Messaoud,$^1$ Thomas Vuillaume,$^1$ and Tom Fran\c cois$^1$}
\affil{$^1$LAPP, Univ. Savoie Mont-Blanc, CNRS, Annecy, France; \email{hana.alimessaoud@lapp.in2p3.fr}}

\paperauthor{Ali Messaoud Hana}{hana.alimessaoud@lapp.in2p3.fr}{}{LAPP, Univ. Savoie Mont-Blanc, CNRS}{}{Annecy}{}{74000}{France}
\paperauthor{Vuillaume Thomas}{thomas.vuillaume@lapp.in2p3.fr}{0000-0002-5686-2078}{LAPP, Univ. Savoie Mont-Blanc, CNRS}{}{Annecy}{}{74000}{France}
\paperauthor{Fran\c cois Tom}{tom.francois@lapp.in2p3.fr}{}{LAPP, Univ. Savoie Mont-Blanc, CNRS}{}{Annecy}{}{74000}{France}



\begin{abstract}
The CTAO (Cherenkov Telescope Array Observatory) is an international observatory currently under construction. With more than sixty telescopes, it will eventually be the largest and most sensitive ground-based gamma-ray observatory.

CTAO studies the high-energy universe by observing gamma rays emitted by violent phenomena (supernovae, black hole environments, etc.). These gamma rays produce an atmospheric shower when entering the atmosphere, which emits faint blue light, observed by CTAO's highly sensitive cameras. The event reconstruction consists of analyzing the images produced by the telescopes to retrieve the physical properties of the incident particle (mainly direction, energy, and type).

A standard method for performing this reconstruction consists of combining traditional image parameter calculations with machine learning algorithms, such as random forests, to estimate the particle's energy and class probability for each telescope. A second step, called stereoscopy, combines these monoscopic reconstructions into a global one using engineered weighted averages.

In this work, we explore the possibility of using Graph Neural Networks (GNNs) as a suitable solution for combining information from each telescope. The "graph" approach aims to link observations from different telescopes, allowing analysis of the shower from multiple angles and producing a stereoscopic reconstruction of the events. We apply GNNs to CTAO-simulated data from the Northern Hemisphere and show that they are a very promising approach to improving event reconstruction, providing a more performant stereoscopic reconstruction. In particular, we observe better energy and angular resolutions(before event selection) and better separation between gamma photons and protons compared to the Random Forest method.

\end{abstract}

\section{Introduction}

CTAO will observe cosmic ray events indirectly by capturing the Cherenkov light produced by their interaction with the atmosphere. 
The event reconstruction is the analysis of the captured images to determine the physical properties of the cosmic ray. 
One step of this reconstruction is the stereoscopy, which is performed after reconstructing the parameters per telescope.
The standard method for stereoscopic reconstruction combines individual telescope reconstructions using weighted averages (for class and energy estimation) and geometric methods for the direction.
This approach is straightforward and has shown consistent, good results \citep{linhoff2023ctapipe}.

In this work, we explore the use of graph neural networks (GNN's) to perform the stereoscopy.

\section{Methods}
\subsection{Graphs}

Objects in everyday life are often defined based on their connections to other objects. 
A graph is a representation of these relationships between entities. It consists of nodes, which represent the entities, and edges, which represent the relationships between them.

In the context of stereoscopic event reconstruction, we may consider each telescope as a node and an array of telescopes observing an event as a graph.

When an atmospheric shower is observed by multiple telescopes, edges are created in the graph to connect the nodes corresponding to the telescopes that observed the same event. Thus, each unique observation of an atmospheric shower captured by an array of telescopes is then represented by a graph.
The telescopes that did not detect any signal are not included in the graph.

Each node will have as features the set of parameters extracted from each telescope observation, called Hillas parameters \citep{Hillas1985} as well as the type of telescope associated with the node (each node has 42 features in total) (see Figure~\ref{fig1}).

\subsection{Dataset}
Our dataset contains 13 telescopes, 4 of which are large-sized telescopes (LST) and 9 of the medium-sized telescopes (MST).
The graphs can have between 2 and 13 nodes.
We used Monte Carlo simulations generated with the CORSIKA software \citep{heck2019corsika}, which simulates atmospheric showers induced by high-energy cosmic rays. The telescope's response to the signal is then simulated using the simtelarray software \citep{bernlohr2008simulation}.The simulations sample used is prod5b\citep{gueta2021cherenkov}.
The data is afterwards analyzed using ctapipe v0.19.3 \citep{linhoff2023ctapipe} up to data level DL2, which includes simulated parameters (ground truth), image parameters, and reconstructed event parameters, used here as the baseline. 
For each telescope, the baseline uses random forests (RF) to predict energies and class probabilities (the gammaness, which represents the confidence level that an event was produced by a gamma ray).
The predictions are then combined using weighted averages. 
The direction is reconstructed through a geometric combination using image parameters.

In this document, we will use the reconstructions obtained by the standard method as a comparison. In this method, the energy and class are reconstructed using weighted averages, while the direction is reconstructed geometrically.

\subsection{Architecture of our graph neural network}

We used the EdgeConv convolution operator introduced by \citep{wang2019dynamic} and implemented in the PyTorch Geometric library.
For each node $n_i$ with features $\mathbf{x}_i$, the operator performs a convolution using the features of $n_i$ and its local neighborhood $\mathcal{N}(i)$. The updated feature $\mathbf{x}'_i$ at node $n_i$ is computed as:
\begin{equation}
   \mathbf{x}'_i = \sum_{j \in \mathcal{N}(i)} MLP \left( \mathbf{x}_i, \mathbf{x}_j - \mathbf{x}_i \right), 
\end{equation}

The architecture consists of 5 EdgeConv layers, each followed by batch normalization and GELU activation. 
In each EdgeConv block, the edge function is implemented as an MLP to compute edge features from the concatenated node embeddings. Global max-pooling and global mean-pooling are applied after all the EdgeConv layers to aggregate node features.
The concatenated result of these global pooling operations is then passed through a final, fully connected layer to produce the final prediction (see Figure~\ref{fig1}). 
In our study, we tested several configurations of the number of EdgeConv layers and found that 5 layers gave the best results. Beyond a certain number of layers, the performance did not necessarily improve.
A Bayesian search was performed to determine the optimal parameters.
The model was trained on a validation dataset and then used for inference on a test dataset.
\articlefigure[width=1\textwidth]{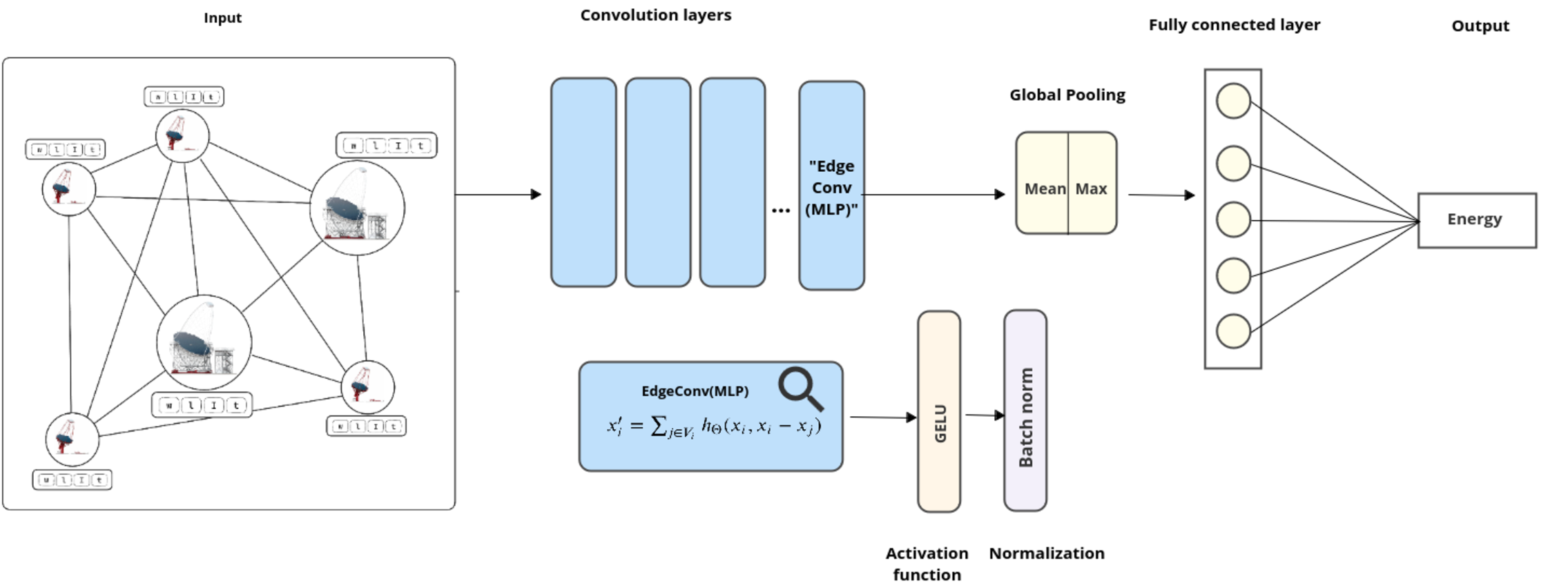}{fig1}{\emph{Right:} Diagram illustrating the architecture of the model. The model takes graphs (produced using a function that connects the graphs  that observed the same event) as inputs, which are processed through multiple EdgeConv convolution layers, followed by batch normalization layers and activation functions (GELU). The features are then aggregated by a pooling layer before passing through a final linear layer. The output dimension varies depending on the specific task: 1D for energy prediction and classification, and 2D for direction prediction (altitude and azimuth).  \emph{Left: }Schematic representation of a graph. Each node of the graph is a telescope of the array. Each node has the Hillas parameters as node features. For each graph, we predict a global graph reconstruction: the direction, energy, and class probability.}

\section{Results}
After training our model, we computed the Instrument Response Functions (IRFs) (see Figure~\ref{fig2})  using pyirf \footnote{\url{https://zenodo.org/records/11190775}} and reconstructed gamma events (all events, or the 30\% most gamma-like per bin of energy based on their reconstructed gammaness). We observe globally better reconstruction performances for all tasks and all energies using GNNs with higher ROC AUC, lower energy, and angular resolutions. However, after event selection, the improvement disappears for the direction reconstruction task, and the reconstruction performance even worsens at energies above 100 GeV, which leaves room for further investigation to improve and optimize the results.

 \articlefigure[width=1\textwidth]{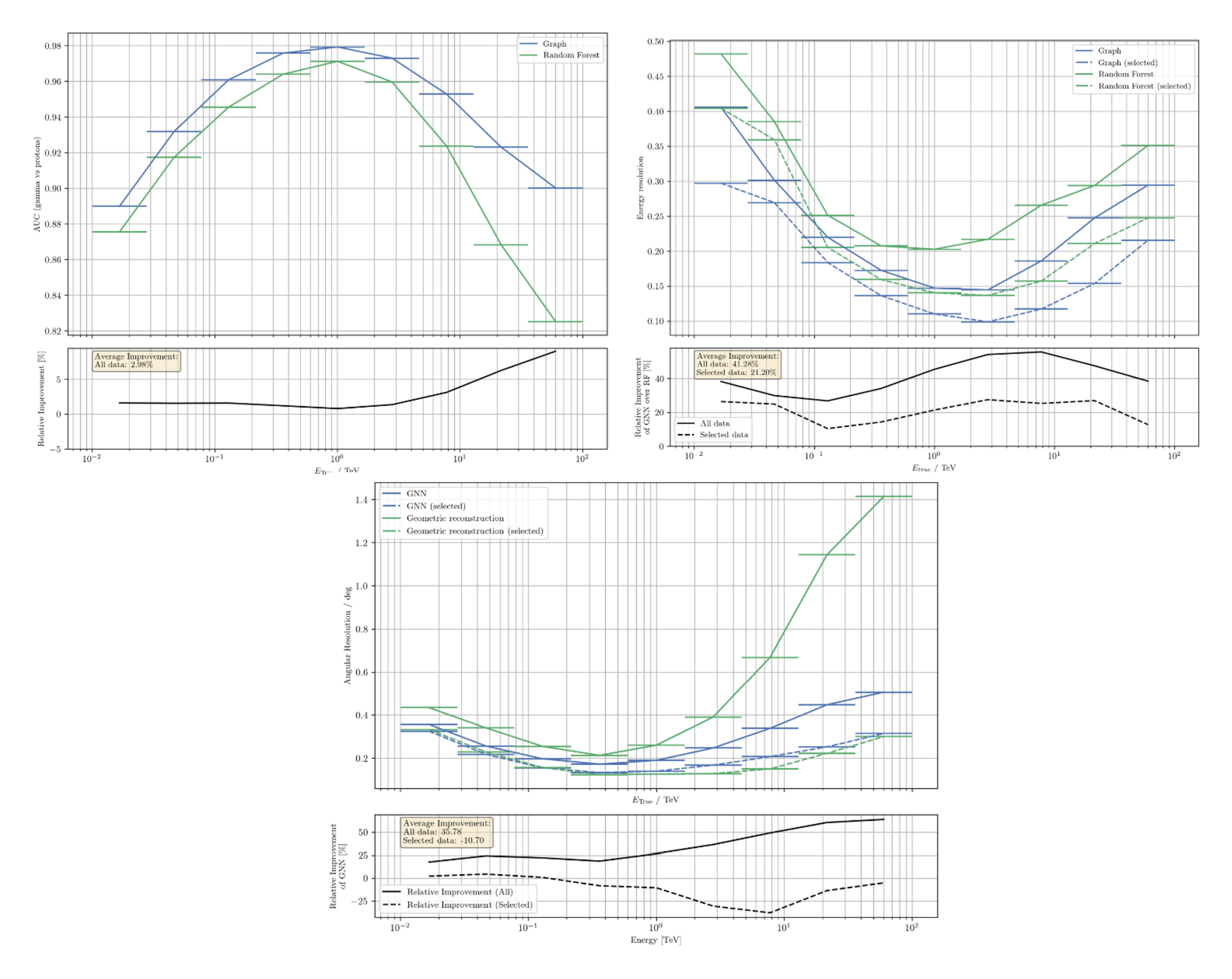}{fig2}{\emph{Top left:} classification performance is improved with higher ROC AUC at all energies. \emph{Top right: }energy resolution is better overall. \emph{Bottom:} angular resolution improves before event selection but deteriorates after event selection}
\newpage
\section{Conclusion}
Graphs show great potential for improving the stereoscopic reconstruction of gamma events.
We observed a significant improvement in the results. However, a major challenge with GNNs is their lack of transparency, making it difficult to understand how they work.
Despite this, GNNs offer a promising approach for the stereoscopic reconstruction of gamma events.


\bibliography{P2-14}  

\end{document}